\documentclass[aps,prl, showpacs,twocolumn,groupedaddress]{revtex4}

\usepackage{amsmath,amssymb}
\usepackage{graphicx}
\usepackage{bm}
\usepackage{ulem}

\begin{document}

\title{dHvA Oscillations in High-Tc Compounds}
\author{L.~Thompson$^{1}$}
\author{P.~C.~E.~Stamp$^{1,2}$}
\affiliation{$^1$Department of Physics and Astronomy, University of
British Columbia, Vancouver, BC, V6T 1Z1, Canada}
\affiliation{$^2$Pacific Institute for Theoretical Physics,
University of British Columbia, Vancouver, BC, V6T 1Z1, Canada}

\date{\today}

\begin{abstract}
Recent de Haas-van Alphen (dHvA) experiments on high-Tc compounds
have been interpreted using Lifshitz-Kosevich (LK) theory, which
ignores many-body effects. However in quasi-2d systems, interactions
plus Landau level quantization give strong singularities in the
self-energy $\Sigma$ and the thermodynamic potential $\Omega$. These
are rapidly suppressed as one increases the c-axis tunneling
amplitude $t_\perp$ and/or impurity scattering. We show that 2d-3d
crossover and interaction effects should show up in these
experiments, and that they can lead to strong deviations from LK
behaviour. Moreover, dHvA experiments in quasi-2d systems should
clearly distinguish between Fermi liquid and non-Fermi liquid
states, for sufficiently weak impurity scattering.
\end{abstract}


\maketitle

By tradition de Haas-van Alphen (dHvA) experiments are interpreted
using Lifshitz-Kosevich (LK) theory, in which magnetization
oscillations probe directly the quasiparticles at the Fermi surface
(so that in a non-Fermi liquid (NFL), with zero quasiparticle weight
on this surface, LK theory implies no dHvA oscillations at all).
Where applicable, LK theory allows unambiguous measurement of Fermi
surface cross-sectional areas, Fermi surface scattering rates, and
Fermi surface band masses \cite{LK}.

Even in 3d, LK theory is not strictly valid because of interactions
\cite{lutt,ES}; these cause ``Engelsberg-Simpson'' (ES) deviations
from LK, which are seen in experiments \cite{expES}. In 2d, the mere
existence of the Fractional Quantum Hall Liquid (FQHL), even when
the interaction strength $\bar V \ll \hbar \omega_c$, shows that
Fermi liquid (FL) theory must break down in a field, provided
impurity scattering is weak \cite{FHL} (ie., once $\omega_c \tau \gg
1$, where $\omega_c$ is the cyclotron frequency and $\tau$ an
impurity scattering time).

Thus the dHvA experiments recently performed in high-Tc systems
\cite{ybco} create a clear paradox. Impurity scattering is weak (it
must be for a dHvA signal to be seen) and the c-axis tunneling
amplitude $t_\perp$ is very small (in YBa$_2$Cu$_3$O$_{7-\delta}$,
$t_\perp \sim$ 15 K is found for $\delta = 0.5$): thus $\hbar
\omega_c > t_{\perp}$ and the system is reaching the 2d limit. And
yet it is claimed that the data can be fit using LK theory
\cite{ybco}. Similar LK analyses have been made for other quasi-2d
systems \cite{SrO,organics}. Since LK theory must break down for
genuinely 2d systems if $\omega_c \tau \gg 1$ and correlations are
strong, this raises several important questions:

(a) How can one generalise dHvA theory to include interactions in
quasi-2d systems; and how should dHvA data then be analysed?

(b) What kind of oscillations will be shown by NFL systems; and can
one tell the difference between FL and NFL states from dHvA
experiments?

To address these questions, we first analyze the 1-particle Green
function $\mathcal G$ and the thermodynamic potential $\Omega$ for a
quasi-2d system, with $t_\perp / \hbar \omega_c$ assumed arbitrary
(but $t_\perp \ll \mu$, the chemical potential). When interactions
are added, we find highly singular behaviour in $\mathcal G$. When
$t_\perp \ll \hbar \omega_c$ and $\omega_c \tau \gg 1$, these
singularities imply a complete breakdown of standard Fermi liquid
theory. However we still find dHvA oscillations, although not of LK
form. To illustrate these results we compute $\mathcal G$ for 2
examples; a NFL with singular forward scattering interactions, and a
FL of band electrons interacting with nearly antiferromagnetic spin
fluctuations. We find clear 3d-2d crossover effects as $\hbar
\omega_c$ exceeds $t_{\perp}$, and departures from LK behaviour
whose form depends strongly on the nature of the many-body
interactions. Neither LK theory, nor its ``ES'' generalisation
\cite{ES}, apply strictly unless $\hbar \omega_c < t_\perp$ and/or
$\omega_c\tau \ll 1$; neither condition is satisfied in experiments.
We find that dHvA experiments ought to be able to distinguish FL
from NFL states.

\vspace{2mm} \noindent {\bf (i) Singularities of $\mathcal G$}: The
form of the dHvA oscillations can be found from either the spectral
function ${\mathcal Im \mathcal G}(\epsilon)$, or directly from
$\Omega$. In 2d, the Landau levels are massively degenerate, and
${\mathcal Im \mathcal G}_\nu(\epsilon) \propto \delta(\epsilon -
\epsilon_\nu)$ where $\epsilon_\nu$ is the $\nu$-th Landau level
energy; interactions destabilize this degeneracy, and so have a
singular effect on ${\mathcal G}(\epsilon)$. However any impurity
scattering or c-axis tunneling tends to suppress this singularity.
Although the analytic structures of ${\mathcal G}(\epsilon)$ and
$\Omega$ are now understood for {\it neutral} 2d fermions
\cite{Fink} in a field (ie. without Landau quantization), there are
no general results when one has both Landau quantization and
interactions \cite{exactFHL}. However, we can derive results for
particular models. Here we discuss 2 simple models involving
quasi-2d band electrons, with dispersion $\epsilon_{\bf k} =
\varepsilon(k_x, k_y) - 2t_\perp\cos(k_za) - \mu$, where $t_{\perp}
\ll \mu$. These couple to low-energy fluctuations; in a finite
field, the lowest-order ``1-fluctuation'' graph for the self-energy
takes the form
\begin{align}
 \label{2ndpert}
\Sigma_\nu(k_z,z) =  & \sum_{\bf q} \sum_{\nu'}\int {d\omega \over
\pi}
|\Lambda_{\nu \nu'}({\bf q})|^2 {\mathcal Im}\chi({\bf q},\omega)\\
& \times \left( {1-f_{\nu'} + n(\omega) \over
z-\omega-\epsilon_{\nu'}(q_z)} + {f_{\nu'} + n(\omega) \over
z+\omega-\epsilon_{\nu'}(q_z)}\right) \nonumber
\end{align}
where $\chi({\bf q}, \omega)$ is the fluctuation propagator,
$f_{\nu} = f(\epsilon_{\nu q_z})$ is the Fermi function for
electrons in the $\nu$-th Landau level, $n(\omega)$ the Bose
function, and the matrix element $\Lambda_{\nu \nu'}({\bf q})$,
between Landau states $\nu, \nu'$ and the fluctuations, incorporates
the fermion-fluctuation coupling $g_q$. When $\mu \gg \omega_c$,
$|\Lambda_{\nu \nu'}({\bf q})|^2 \sim g_q^2 (m/2 \mu)^{1\over2}
\omega_c / \pi q$.

At this time there is no consensus on a model for high-$T_c$
superconductors (indeed the central issue is whether they are FL or
NFL); and other strongly-correlated quasi-2d systems are quite
complex. Thus, instead of presenting numerical calculations for a
specific experimental system, we address the general questions posed
in the introduction by analysing two widely studied models of strong
correlations in quasi-2d systems: in zero field these describe a FL
and NFL respectively.

We begin by discussing the self-energy, which for a quasi-2d system
can be written near the Fermi surface as $\Sigma(z) = \bar \Sigma(z)
+ \Sigma_{osc}(z)$, where $\bar \Sigma (z)$ is non-oscillatory in
$1/B$, and the oscillatory part
\begin{equation}
 \label{Sosc}
\Sigma_{osc} (z) = 2 \sum_{r=1}^{\infty} (-1)^r  \Sigma_r(z) J_0
\left( 4\pi r {t_\perp \over \hbar \omega_c} \right) \sin\left(2\pi
r {A_F \over B} \right)
\end{equation}
The Bessel function $J_0$ in this expression comes from integrating
over $q_z$.

{\it Model (a) Spin fluctuation model}: This well-known model
\cite{nAFM} has 2d lattice fermions with dispersion
\begin{equation}
\varepsilon(k_x,k_y) = -2t_0 (\cos k_x+ \cos k_y ) -
 4t_1\cos k_x\cos k_y
\end{equation}
and coupling $t_{\perp}$ between planes; the fermions couple to
antiferromagnetic spin fluctuations, with propagator
\begin{equation}\label{spinsusc}
\chi({\bf q},\omega) = {\chi_{0} \over 1+\xi^2({\bf q}-{\bf Q})^2 -
i\omega/\omega_{SF}}
\end{equation}
via a coupling $g_q = g$. The wave-vectors ${\bf Q} = (\pm \pi, \pm
\pi)$. In zero field this model, with or without vertex corrections
\cite{vertexcor}, gives FL behaviour, with a Green function having
finite residue $z_{k_F}(\mu)$ at the Fermi surface, and a
self-energy $\Sigma(\omega)$ with a 2d FL form (ie., with ${\mathcal
Re} \Sigma(\omega) = (1 - m/m^*)\omega$ and ${\mathcal
Im}\Sigma(\omega) \propto \omega^2(1 + \ln \omega)$).

\begin{figure}[t]
\begin{center}
\includegraphics[width=7.5cm]{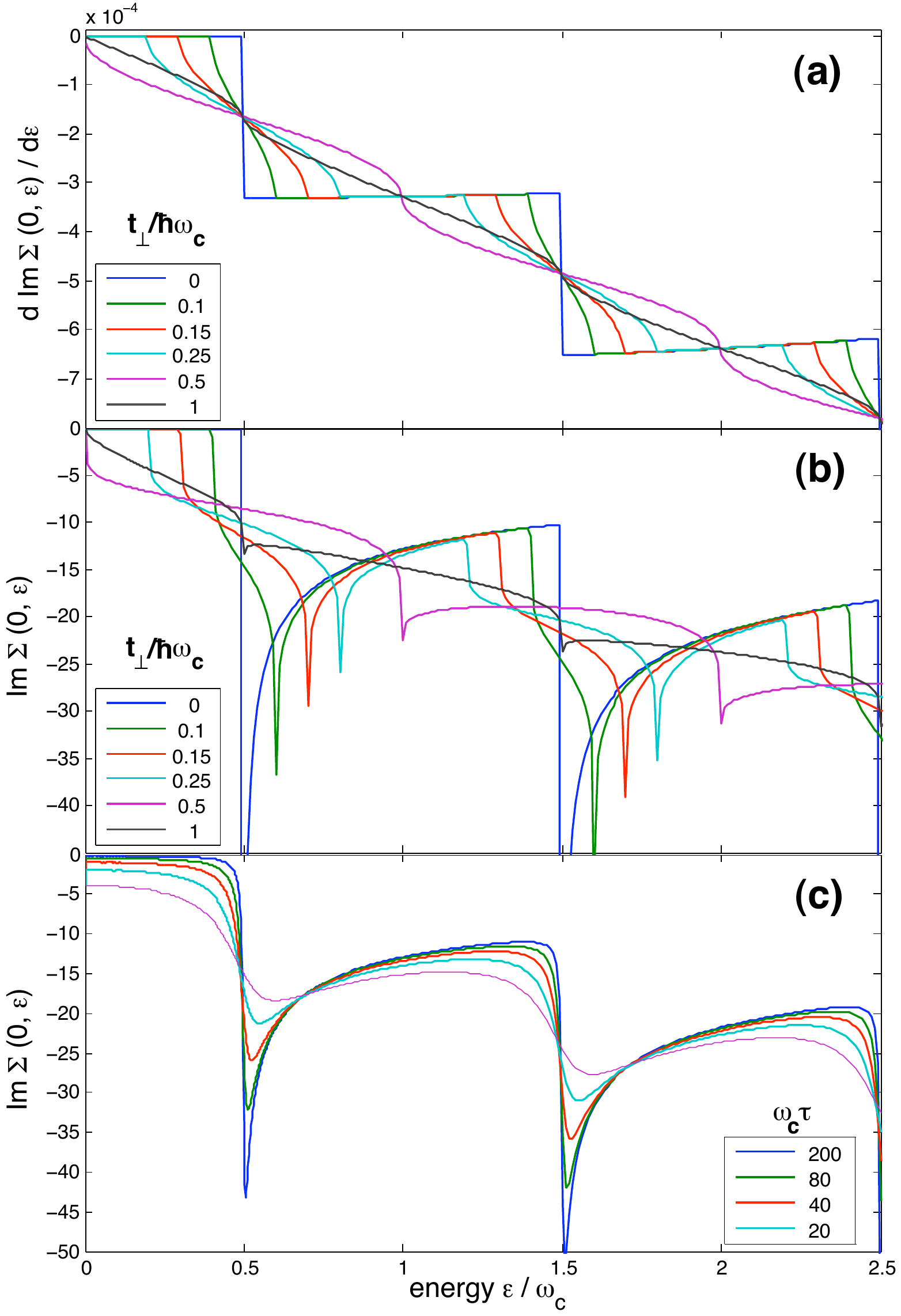}
\caption{The imaginary part ${\mathcal Im}\Sigma(\epsilon)$ of the
self-energy, as a function of $t_{\perp}/\hbar \omega_c$, at $T=0$.
(a) shows ${\mathcal Im}\partial \Sigma/\partial \epsilon$ for the
spin fluctuation model; we fix $g_q = g = 0.58$ eV, $\chi_0 =
80$ states/eV, $\xi = 2.5a$ and $\omega_{SF}=10$ meV. (b) shows
${\mathcal I}m \Sigma$ (with no derivative) for the non-Fermi liquid
model, assuming $s=3$; we fix $K_{s=3} = 0.013
(\hbar\omega_c)^{2/3}\mu^{1/3}$ with $\mu = 6000~K$. (c) shows the
effect of impurity scattering on the NFL model; we plot ${\mathcal
Im}\Sigma(\epsilon)$ for different values of $\omega_c \tau$,
assuming $t_{\perp}/\hbar\omega_c =  0.75$. }
 \label{ImSig}
\end{center}
\end{figure}

In a finite field, $\Sigma(\omega)$ can be evaluated analytically,
but the expression is extremely lengthy \cite{Sigrs}. The essential
result is shown in Fig. \ref{ImSig}; Landau quantization introduces
a ``step-like'' behaviour in $\partial {\mathcal
Im}\Sigma/\partial{\epsilon}$, with corresponding singularities in
${\mathcal Re}\Sigma(\epsilon)$, at $\epsilon = \epsilon_{\nu}$.
Notice how rapidly this singular behaviour is suppressed by
interplane hopping -- it is almost invisible once $t_{\perp} \sim
\hbar \omega_c$. Impurity scattering has a similar effect (not shown
in Fig. \ref{ImSig}).

{\it Model (b) Non-Fermi liquid model}: We now couple the band
electrons to fluctuations with propagator \cite{gauge}:
\begin{equation}
 \label{singchi}
\chi({\bf q},\omega) = {q \over \chi q^s - i \gamma \omega}
\end{equation}
where $s$ is a dynamic scaling exponent, with $2\le s \le 3$, using
a fermion fluctuation coupling $g_q = K_s$. The zero field
self-energy has the NFL forms $\Sigma(\epsilon) \sim \epsilon \ln
\epsilon$ (for $s=2$) and $\Sigma(\epsilon) \sim
(i\Omega_0/\epsilon)^{1\over3} \epsilon$ (for $s=3$), so that
$z_{\bf k}(\epsilon) \to 0$ on the Fermi surface. In a finite field,
$\Sigma_\nu(k_z,z)$ can again be evaluated analytically in the form
(\ref{Sosc}), with the coefficients $\Sigma_r(z)$ taking the
interesting $T=0$ form :
\begin{align}
\Sigma_r(z) \;=\; & {sK_s \over 2 r^{2/s}}
 \left[ {\cal Z}_r^{1\over2} S_2\left( {1 \over 2} + {2 \over s},{1 \over 2}; {\cal Z}_r \right) - {\cal Z}_r^{2/s} \right. \\
 & -\;\left. (-{\cal Z}_r)^{1\over2} S_2\left( {1 \over 2} + {2 \over s},{1 \over 2}; -{\cal Z}_r \right) + (-{\cal Z}_r)^{2/s}
 \right]
 \nonumber
\end{align}
(and a more complicated finite $T$ form), where $S_2$ is a Lommel
function \cite{Lommel}, and ${\cal Z}_r = 2\pi r z/\hbar \omega_c$.
Now the singular behaviour in $\Sigma$ is far more pronounced;
again, it is eliminated by switching on $t_\perp$ (Fig.
\ref{ImSig}(b)), or by impurity scattering (calculated in Fig.
\ref{ImSig}(c) in a self-consistent Born approximation).

We see that both models show singular behaviour of
$\Sigma(\epsilon)$ as a function of $\epsilon$, implying similar
behaviour for the quasiparticle weight $z_k(\epsilon)$. At the Fermi
energy, $\Sigma(\epsilon=\mu)$ will then show the same singular
behaviour as a function of $B$, periodic in $1/B$. Strictly
speaking, this means a breakdown of FL theory for both models, but
much more strongly for the NFL system. Because these singularities
are rapidly suppressed by both inter-plane hopping and impurity
scattering, this breakdown will only be clearly visible when
$t_{\perp}, \hbar/\tau \ll \hbar \omega_c$.

\vspace{3mm}

{\bf (ii) Thermodynamic potential $\Omega$}: If ``crossed graphs''
can be ignored in $\Sigma(z)$, we can write an expression for
$\Omega$ in terms of ${\cal G}$ \cite{CS}:
\begin{equation}\label{Omega}
\Omega = -{1\over \beta} {\mathrm Tr} \ln \left[ \left( \bar
{\mathcal G} + {\mathcal G}_{osc} \right)^{-1} \right]
\end{equation}
where $ \bar {\mathcal G}$ is the non-oscillatory part of ${\mathcal
G}$. This expression resembles the classic Luttinger/ES expression
\cite{ES} for $\Omega$, except that the latter drops ${\mathcal
G}_{osc}$ from (\ref{Omega}). This is justified in 3d, but not in 2d
\cite{CS}; in the quasi-2d case it is only justified if $t_\perp \gg
\hbar \omega_c$. From (\ref{Omega}) we find $\Omega = \bar \Omega +
2 \sum_{r=1}^\infty (-1)^r \Omega_r \cos ( r\hbar A_F/ eB )$, where
$\bar \Omega$ is the non-oscillatory part of $\Omega$, and
\begin{align}
 \label{Omegar}
\Omega_r  = \;- &{m\over \hbar^2 \beta} \sum_{n>0} \;[ J_o^2\left({4
\pi r t_{\perp} \over \hbar \omega_c}\right)
\zeta_r(\omega_n)  \\
 &- \;{\hbar\omega_c \over 2\pi r} J_o\left({4 \pi r t_{\perp} \over \hbar \omega_c}\right)
 \exp (-{2\pi r\over \hbar\omega_c}
 (\omega_n + \zeta(\omega_n)) ]    \nonumber
\end{align}
where the $\zeta_r(\omega_n) = i \Sigma_r (i\omega_n)$ are real and
positive. Equation (\ref{Omegar}) reduces to the Luttinger/ES
expression for $\Omega$ if we drop the first term, and if in the
second term we use only the non-oscillatory part
$\bar{\zeta}(\omega_n)$ of $\zeta(\omega_n) = i\Sigma (i\omega_n)$.
It further reduces to LK if we assume $\Sigma(\omega) \rightarrow
(1-m/m^*) \omega + i/2\tau$, ie., a mass renormalisation and
scattering rate both independent of energy. Clearly (Fig.
\ref{ImSig}) the oscillatory part of $\Sigma$ must not in general be
neglected.

\begin{figure}[t]
\begin{center}
\includegraphics[width=7.5cm]{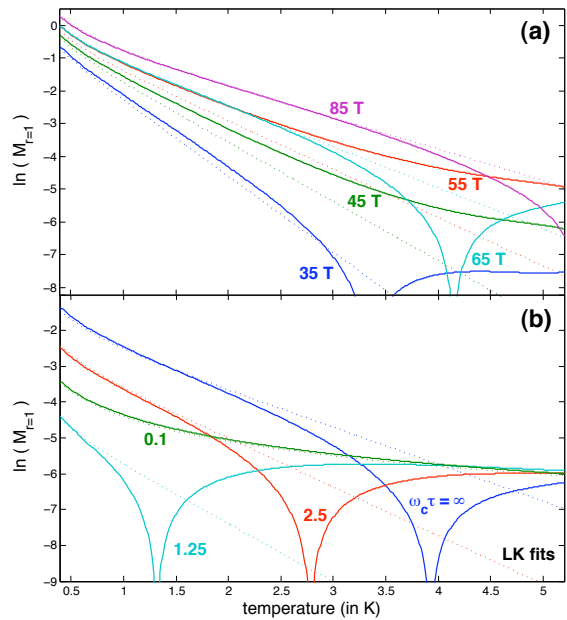}
 \caption{$\ln {M_r}$ for $r=1$ against $T$ (the ``mass plot''),
for the spin fluctuation model: (a) Plot for different fields,
assuming $t_\perp = 15 K$ but no impurity scattering (when $B=55~T$,
$\hbar \omega_c = 20~K$); (b) Plot for different impurity
scattering rates, assuming $t_\perp/\hbar\omega_c = 0.6$. The dashed
lines are fits to LK.}
 \label{massSF}
\end{center}
\end{figure}

\vspace{2mm}

{\bf (iii) Oscillatory Magnetisation}: We write the magnetisation at
constant chemical potential $M_{\mu}(B)  = -\left.\partial \Omega /
\partial B\right|_\mu$ in the form $M = \bar M(B) + 2 \sum_r (-1)^r M_r$,
where $\bar{M}$ is the non-oscillatory part ($M(B)$ at constant $N$
is found by making a Legendre transform \cite{Legendre}).
Differentiating (\ref{Omegar}), we get $M_r = M_1^r + M_2^r$, with
\begin{align}
 \label{M-r}
M_1^r \;=\; &-\Omega_r \sin ( r\hbar A_F/ eB ) \\
M_2^r \;=\; &-{\partial \Omega_r \over \partial B} \cos ( r\hbar
A_F/ eB ) \nonumber
\end{align}
The key point here is that if $\Sigma$ contains strong oscillations
with energy, these translate into a very strong new oscillatory
contribution to $M_{osc}$.

Equation (\ref{M-r}) yields a very rich variety of forms for $M(B)$,
depending on the two parameters $t_{\perp}/\hbar \omega_c$,
$\omega_c \tau$, and on the form and strength of the interactions.
We have no space here to discuss the whole parameter range, but we
can summarize the key features:

\begin{figure}[bt]
\begin{center}
\includegraphics[width=7.5cm]{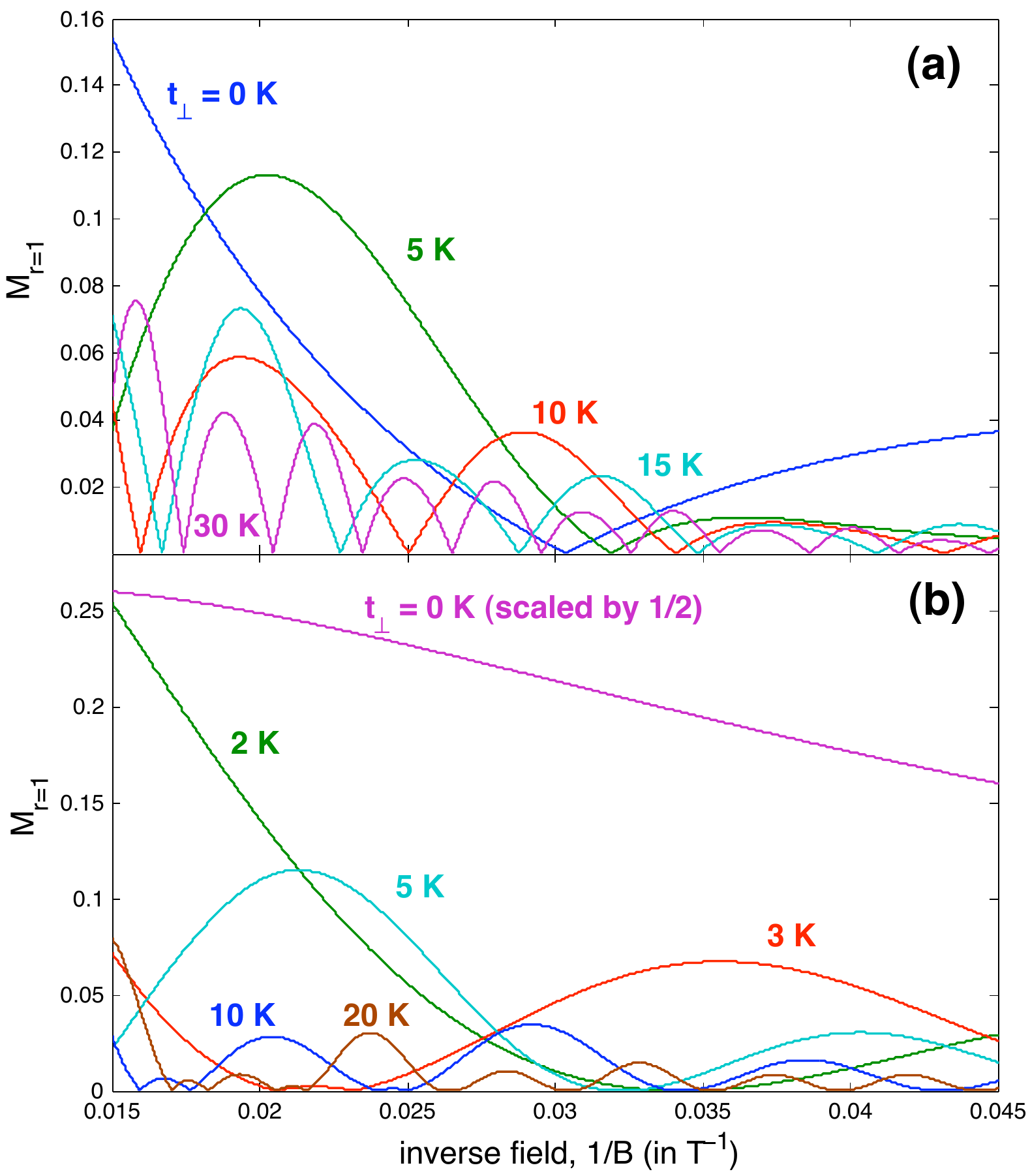}
\caption{The dHvA component $M_r$ for $r=1$, against $1/B$ (again,
if $B=55~T$, $\hbar \omega_c = 20~K$), for different values of
$t_{\perp}$ (measured in $K$), with no impurity scattering; we
assume $T=1~K$. (a) for the spin fluctuation model; and (b) for the
non-Fermi liquid model with $s=3$ (the topmost curve in (b) has been
rescaled by a factor $0.5$). } \label{Mtperp}
\end{center}
\end{figure}

1. Clear departures are seen from LK theory, even in mass plots
(Fig. \ref{massSF}), for NFL systems and even for FL unless the
fluctuation energy scale $\omega_{SF} \gg \hbar \omega_c$, and/or
$\omega_c \tau < 1$. Without interactions, only the 2nd term ($\propto
J_0$) survives in (\ref{Omegar}); this term is well-known in LK theory \cite{J0}. With
interactions, the two terms compete -- as the field-induced
singularities in $\Sigma(\epsilon)$ (and hence in $\zeta(\omega_n)$
become stronger, the 1st term in $\Omega_r$ ($\propto J_0^2$) increases, and
 for NFL it can dominate the 2nd term. The much
stronger singular structure in $\Sigma(\epsilon)$ means that NFL
have much stronger departures from LK than FL.

2. The form of $M(B)$ depends strongly on $t_{\perp}/\hbar
\omega_c$. This gives a remarkable structure in field plots (Fig.
\ref{Mtperp}), which is eliminated by strong impurity scattering
(Fig. \ref{massSF}(b)), or by removing strong correlation effects.

3. Short-range impurity scattering strongly suppresses the singular
structure from interactions once $\omega_c \tau < 1$ (see Fig.
\ref{massSF}(b)). However, curiously, it affects $M_1^r$ and $M_2^r$
rather differently; $M_1^r$ decreases exponentially with $1/\omega_c
\tau$ (\`a la Dingle) but $M_2^r$ decreases approximately as a power
law. More refined analysis of the effect of scattering off
impurities and off small angle scatterers (like dislocations) is
certainly necessary for this problem.

We see that interactions have profound effects on the quasiparticles
and the thermodynamics of conducting systems in high fields, for
quasi-2d systems. These effects are rapidly removed by interplane
coupling (once $t_\perp > \hbar \omega_c$), and even more rapidly by
impurity scattering (once $\omega_c \tau < 1$). The models we have
used are of course rather simple (although very widely used in the
literature); but our main results are not crucially changed by, eg.,
adding vertex corrections.

Consider now the experimental situation. Experiments on YBCO fall
precisely in the crossover regime, with $t_\perp \sim 15$ K, and 15
K $\lesssim \hbar \omega_c \lesssim 30$ K. It is not yet possible to
compare the experimental fits \cite{ybco} on YBCO and Tl-2201 with
the theory here, because these fits have not included the $J_0$ term
(which already exists in LK theory \cite{J0}). It will be extremely
interesting to have fits to different strong-correlation models --
and to discriminate between FL and NFL models. We note that absence
of the $J_0^2$ term in (\ref{Omegar}) would indicate the underlying
state is FL (but NFL if the $J_0^2$ term is strong). It will also be
interesting to look more closely at other strongly-correlated
quasi-2d systems in high fields -- where few departures from LK have
been found so far. Finally, note that any experiments sensitive to
the singular structure we find in ${\cal G}$ should show interesting
effects. Obvious examples are c-axis tunneling and SdH experiments
in very high fields, but a generalisation of the foregoing to a
transport theory will be required.

This work was supported by NSERC, CIFAR and PITP. We thank P. W.
Anderson, D. Bonn, S. Julian, B. Ramshaw, and G. A. Sawatzky for
discussions.


\begin{thebibliography}{99}
\bibitem{vertexcor} Vertex corrections to eqn (\ref{2ndpert}) change
        the numerical results, but not the basic form of ${\mathcal Im} \Sigma$,
        or of the dHvA oscillations, for the spin fluctuation model.
\bibitem{Sigrs} The self-energy $\Sigma(z)$, and the coefficients $\Sigma_r(z)$,
        can be found analytically for the spin fluctuation model as a
        a rather cumbersome function of elliptic integrals and exponential integrals.
\bibitem{Legendre} It is convenient to do field-theoretical
calculations at constant $\mu$. To find, eg., $M_N(B)$ (ie., $M$ at
constant $N$), we calculate $F = \Omega + \mu N$, with $N = - Tr
{\mathcal Im}{\cal G}/\pi$; then $M_N = -\partial F/\partial B$.
\end{thebibliography}
\end{document}